# Systematic and Integrative Analysis of Proteomic Data using Bioinformatics Tools

Rashmi Rameshwari
Asst. Professor, Dept. of Biotechnology,
Manav Rachna International University,
Faridabad, India

Dr. T. V. Prasad
Dean (R&D), Lingaya's University,
Faridabad, India

***Abstract*— The analysis and interpretation of relationships between biological molecules is done with the help of networks. Networks are used ubiquitously throughout biology to represent the relationships between genes and gene products. Network models have facilitated a shift from the study of evolutionary conservation between individual gene and gene products towards the study of conservation at the level of pathways and complexes. Recent work has revealed much about chemical reactions inside hundreds of organisms as well as universal characteristics of metabolic networks, which shed light on the evolution of the networks. However, characteristics of individual metabolites have been neglected in this network. The current paper provides an overview of bioinformatics software used in visualization of biological networks using proteomic data, their main functions and limitations of the software.**

***Keywords- Metabolic network; protein interaction network; visualization tools.***

## I. Introduction

Molecular interaction network visualization is one of the most user-friendly features developed for the simulation process of biological interactions [29]. Drawing of any molecule for example, protein may seem to be easy but generating the same protein with all the types of conformation that it can attain during any interactions and to simulate this process is quite difficult. In this context one of the greatest manually produced molecular structures of its time was done by Kurt Kohn's 1999 map of cell cycle control. Protein interactions network visualization deals with territory that is very similar to that of protein-protein interaction prediction, but differs in several key ways. proteomics data are often associated with pathways or protein interactions, and both of these are easily visualized as networks[22]. Even types of data not normally viewed as networks (*e.g.* microarray results) are often painted onto signaling, metabolic, or other pathways or protein interaction networks for visualization and analysis.

Visualization and analysis tools are commonly used to interact with proteomic data. Most, visualization tools were developed simply for illustrating the big picture represented by protein-protein interaction data or expression data, for qualitative assessment, not necessarily for quantitative analysis or prediction. Expression and interaction experiments tend to be on such a large scale that it is difficult to analyze them, or indeed grasp the meaning of the results of any analysis. Visual representation of such large and scattered quantities of data allows trends that are difficult to pinpoint numerically to stand out and provide insight into specific avenues of molecular functions and interactions that may be worth exploring first out of the bunch, either through confirmation or rejection and then later of significance or insignificance to the research problem at hand. With a few recent exceptions, visualization tools were not designed with the intent of being used for analysis so much as to show the workings of a molecular system more clearly.

Visualization tools also do not actually predict molecular interactions themselves or their characteristics. On the contrary, visualization tools only create a graphical representation of what is already "known" in literature and in molecular interaction repositories such as the Gene Ontology (GO) [7]. A side-effect of displaying interaction networks by treating some proteins as members of more general family groupings or going by interactions in different tissues or as analogues in other tissues or organisms is the apparent display of a protein-protein or other molecular interaction that is inferred but may or may not have actually been observed and documented, something that could be misconstrued as a predicted interaction.

The most significant difference between molecular interaction network visualization and molecular interaction prediction is the nature of the information it provides [11]. Protein-protein interaction prediction is characterized by its concern with how proteins will interact, where they will interact, under what conditions they were interact, and what parts are necessary for their interaction. These characteristics are governed by physical and chemical properties of the proteins or other molecules involved, which may be actual molecules that have been described through extensive proteomics experiments or hypothetical, *in silico* generated species that are being investigated for pharmaceutical and other applications. Interaction network visualization tools are given no knowledge of physical or chemical properties of the proteins that, why they interact. As a result, the information they inadvertently impart concerns only whether or not certain proteins putatively interact with certain other proteins, not how they interact, when they interact, or why they interact.

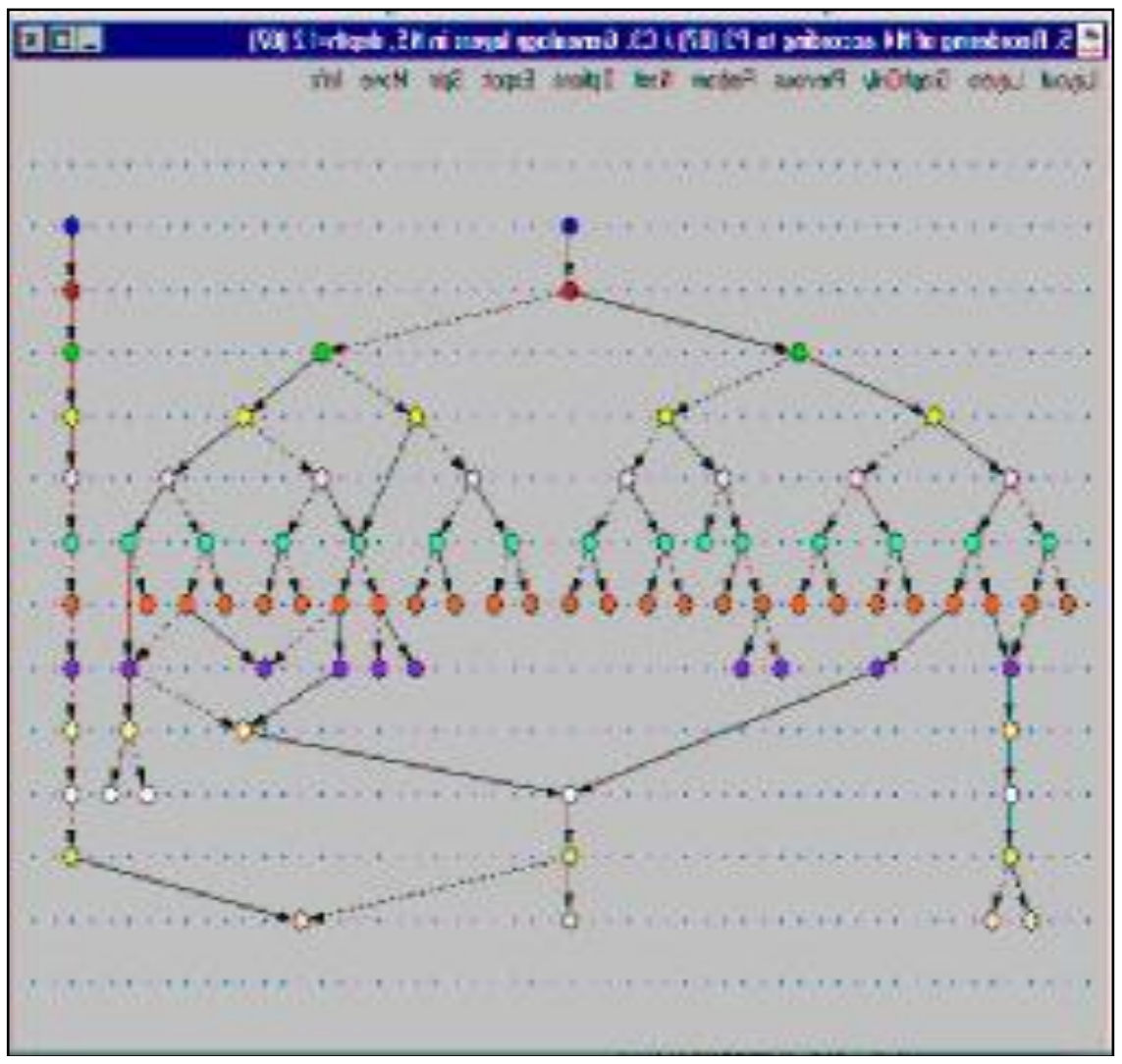

Figure:(A)

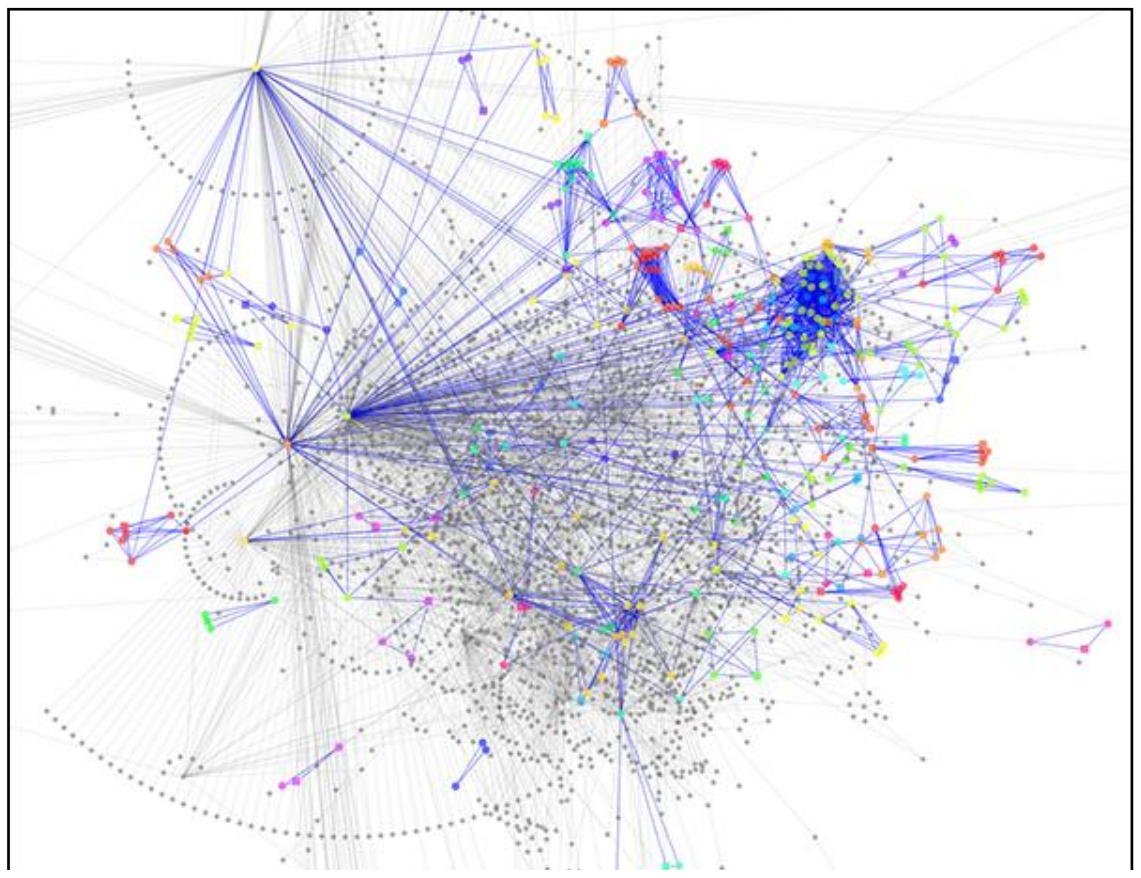

Figure: (B)
Figure1: Comparative visualization of Protein-Network drawn by different tools like Pajek (Fig.A), and Cytoscape (Fig. B). Here genes are represented as nodes and interaction as edges.

## II. Background and Related Work

With the recent advances in high-throughput technologies, software tools have been developed to visualize and analyze large-scale data. This paper deals with various visualization techniques for proteomic data. Major emphasis is on network graph generated during protein-protein interaction. Many tools are being used for this purpose which is based on different algorithm. For example, Pajek [6] and Cytoscape [8] use force directed layout algorithm which produces graph by computing force between pairs of nodes in each iteration of the optimization process. The networks can be visualized as indicated in Fig 1A and Fig 1B.

As protein interactions data also helps in study related to evolutionary analysis. In a new era, it is necessary to understand how the components which involve in biological systems from the various biological data and knowledge of components at molecular level. It reveals the structure of biological systems and lead to “ontological” comprehension of biological systems. Comprehensive data of protein interactions is also suitable for systems level evolutionary analysis.

There are many commercial software such as Ingenuity Pathway Analysis (Figure 2), MetaCore and Pathway studio, designed to visualize high-throughput data in the context of biological networks. Biological networks have a scale-free and modular organization. In a scale free networks the degree of distribution follows a power-law, which means that only a small number of nodes, called hubs are highly connected [12]. Hubs usually play essential roles in biological systems [13]. On the other hand, groups of proteins with similar functions tend to form clusters or modules in the network architecture. Many commercial software for network visualization follow this law.

Metacore is an integrated knowledge database and software suite for pathway analysis of experimental data and gene lists. It is based on manually curetted databases of human protein-protein, Protein-DNA and protein compound interactions. This package includes easy to use, intuitive tools for search, data visualization, mapping and exchange, biological networks and interactomes [32].

The other software tool known as Pathway studio, is based on a mammalian database, named: ResNet 5 mammalian, which is generated by text mining of the PubMed database and 43 full text journals [14]. The advantages of using this tool are that it increases the depth of analysis of high-throughput data generation experiments like microarray gene expression, proteomics, metabolomics. Enables data sharing in a common analysis environment. This tool simplifies keeping up to date with the literature and brings this knowledge into an analysis environment. This also enables visualization of gene expression values and status in the context of protein interaction networks and pathways. However, free software like Pathway Voyager (Figure 5), GenMapp and Cytoscape are also available. Pathway Voyager applies flexible approach that uses the KEGG database [27] to pathway mapping [17]. GenMapp (Figure 6) is also designed to visualize gene expression data on maps representing biological pathways and gene groupings. GenMapp has more option which can modify or design new pathways and apply complex criteria for viewing gene expression data on pathways [14].

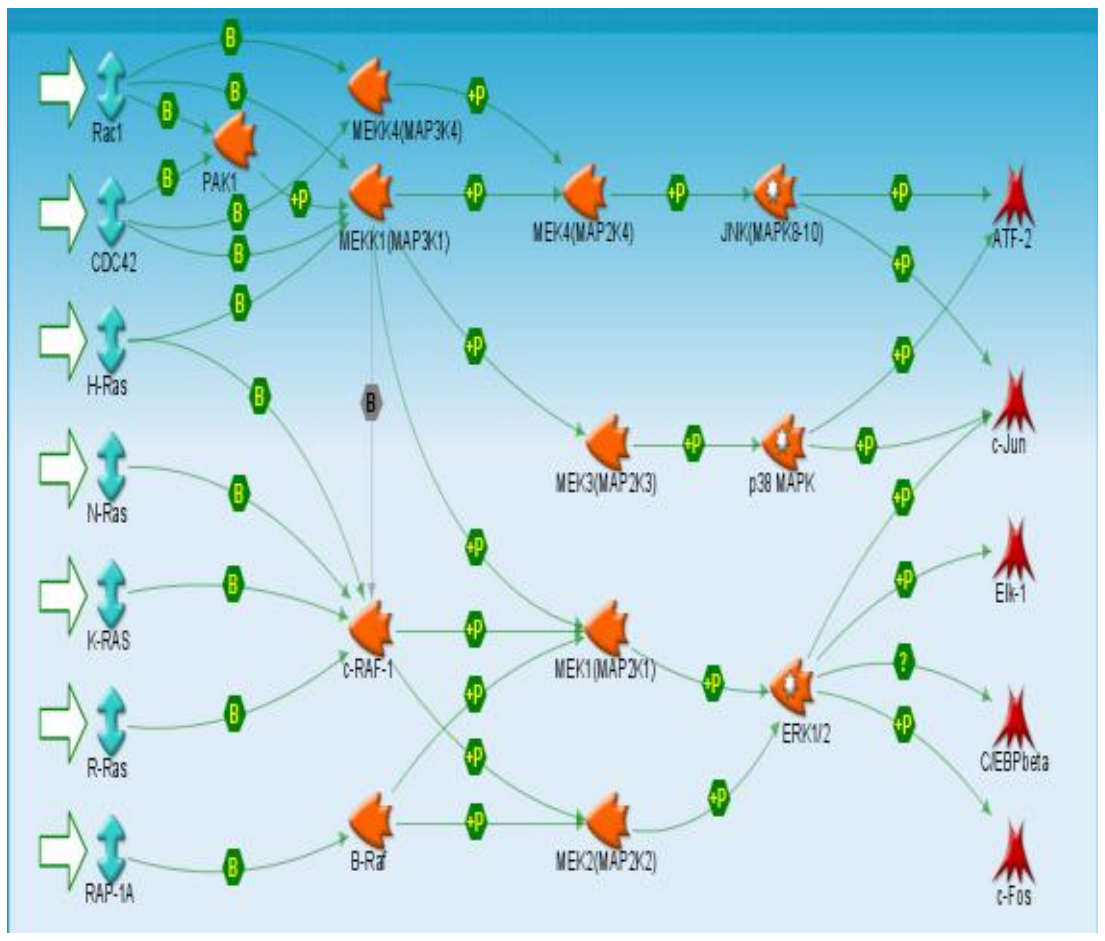

Figure 2: G-Protein signaling_Ras family GTPases in kinase cascades. Image generated by Ingenuity Pathway Analysis.

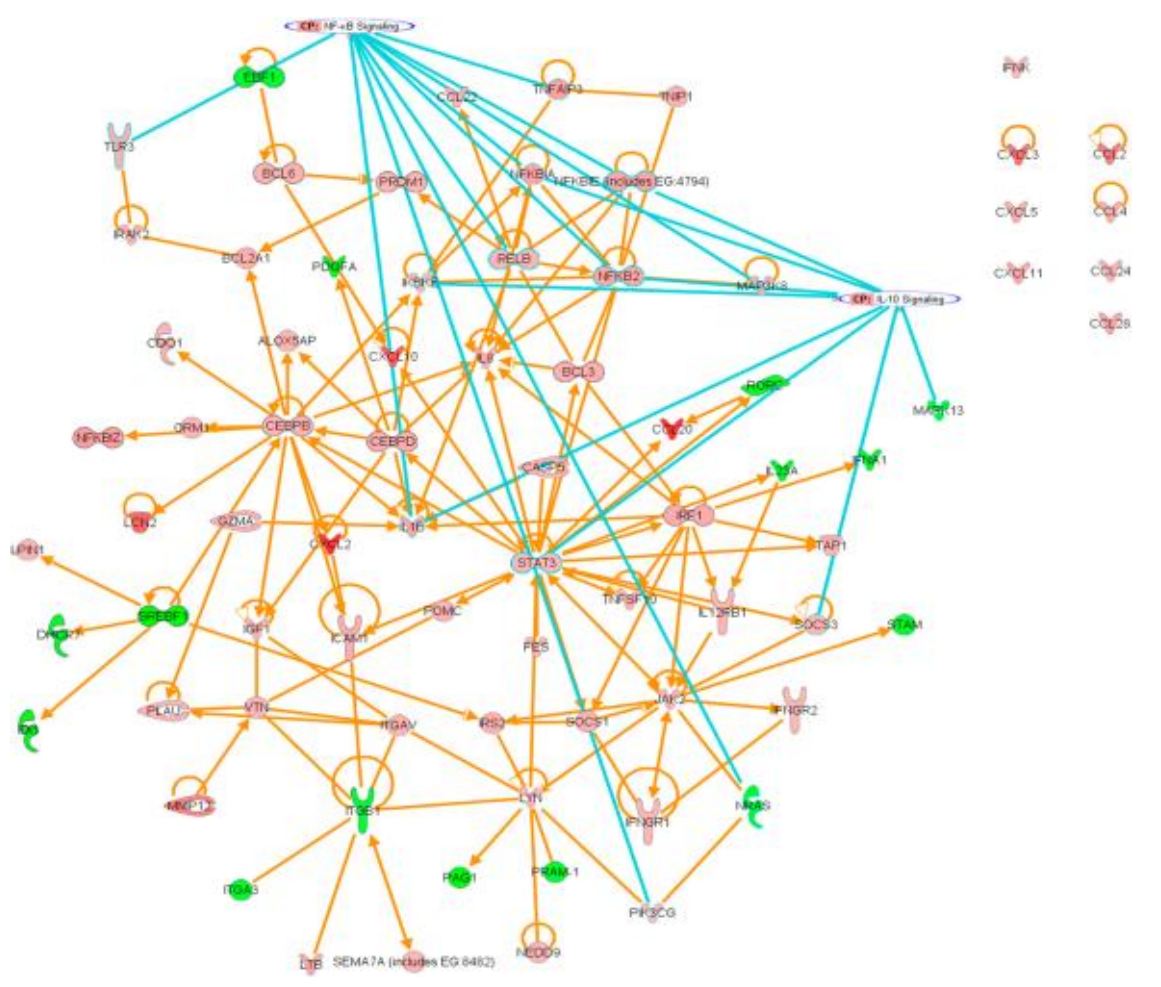

Figure 3: A Network generated by Metacore Software [32]

## III. Limitations and Future Developments

After the completion of Human Genome Project, 36,000 genes were discovered, which has potential to synthesize more than 1,00,000 proteins [19], less than 50% of these genes can be assigned a putative biological function on the basis of sequence data [18]. With the advancement of technology many software has been designed to explore biological networks, like protein interactions network, protein-DNA interactions etc., that are based on databases like DIP [31], MINT [30], of mammalian database. The tools represented in this paper are applicable to a wide range of problems and their distinct features make them suitable for a wide range of applications.

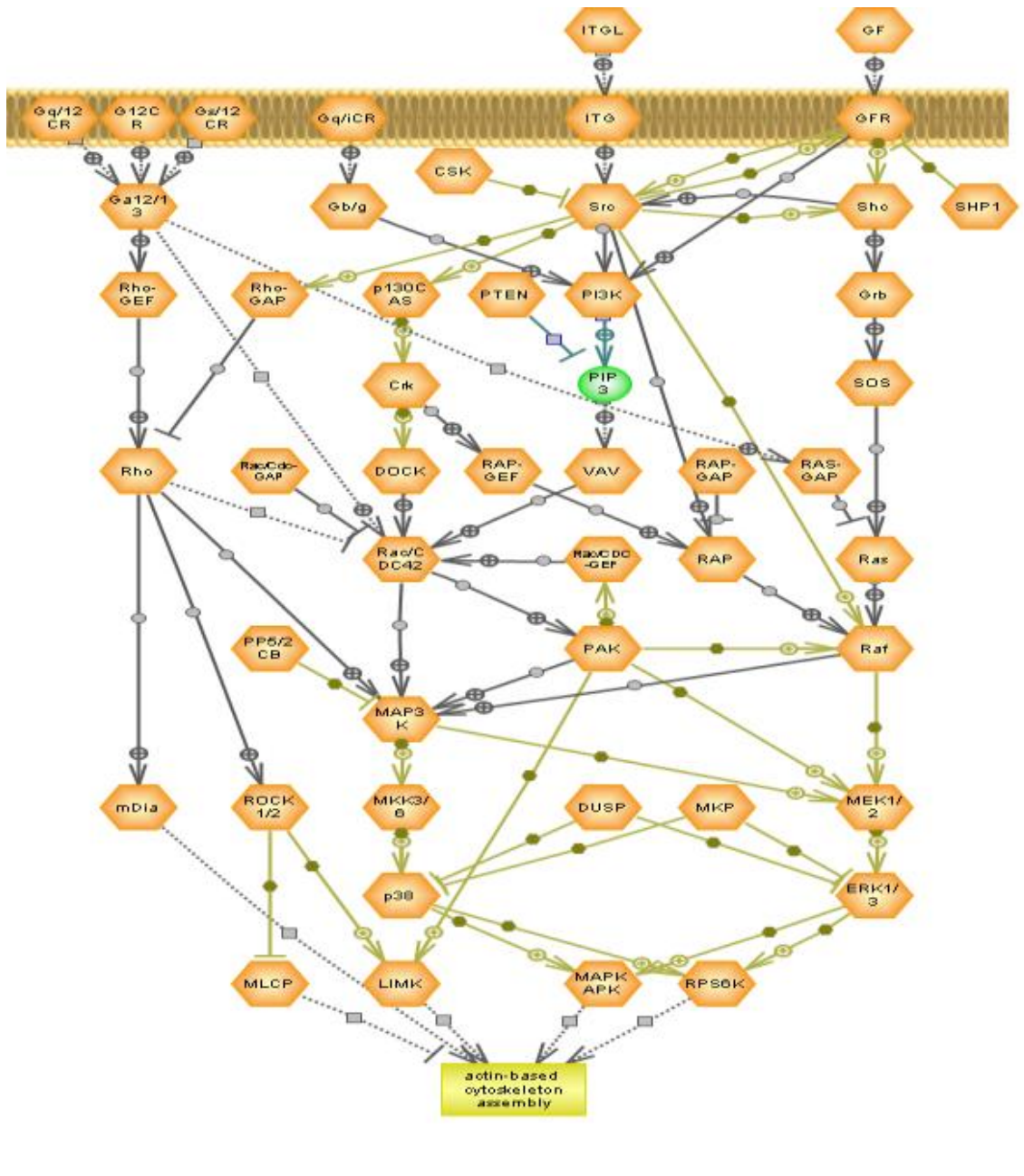

Figure 4: Actin Cytoskeleton Regulation, network generated by Pathway Studio

Clustered graphs are common occurrences in the biological field. Examples of pre-clustered graphs include clustering of proteins or genes based on some biological functionality, structural geometry, expression pattern or other chemical property [11].

In any “post- omic” analysis two essential concepts must be applied to understand biological functions at a systems level. First integrate different levels of information and second, view cells in terms of their underlying network structure. The information about biological entity is scattered in different databases [28]. Hence, the information retrieval from diverse databases is done, which is bit time consuming. Current databases are good for the analysis of a particular protein or small interaction networks. But they are not as useful for integration of complex information on cellular regulation pathways, networks, cellular roles and clinical data and they lack coordination and the ability to exchange information between multiple data sources [20]. There is need of software that can integrate information from different database as well as from diverse sources.

To analyze data, at present many software are there like Ingenuity Pathway analysis, Metacore and Pathway Studio. They work on owner curetted database at the same time there high price make them unaffordable for academic institute to use them. At the same time Cytoscape which has many properties for visualization of high throughput data, can be alternative for users [15].

However network constructed with Cytoscape are sometimes liable to show errors. So there is need to improve quality of available curetted databases and also to develop integrative knowledge bases that are especially designed to construct biological networks.

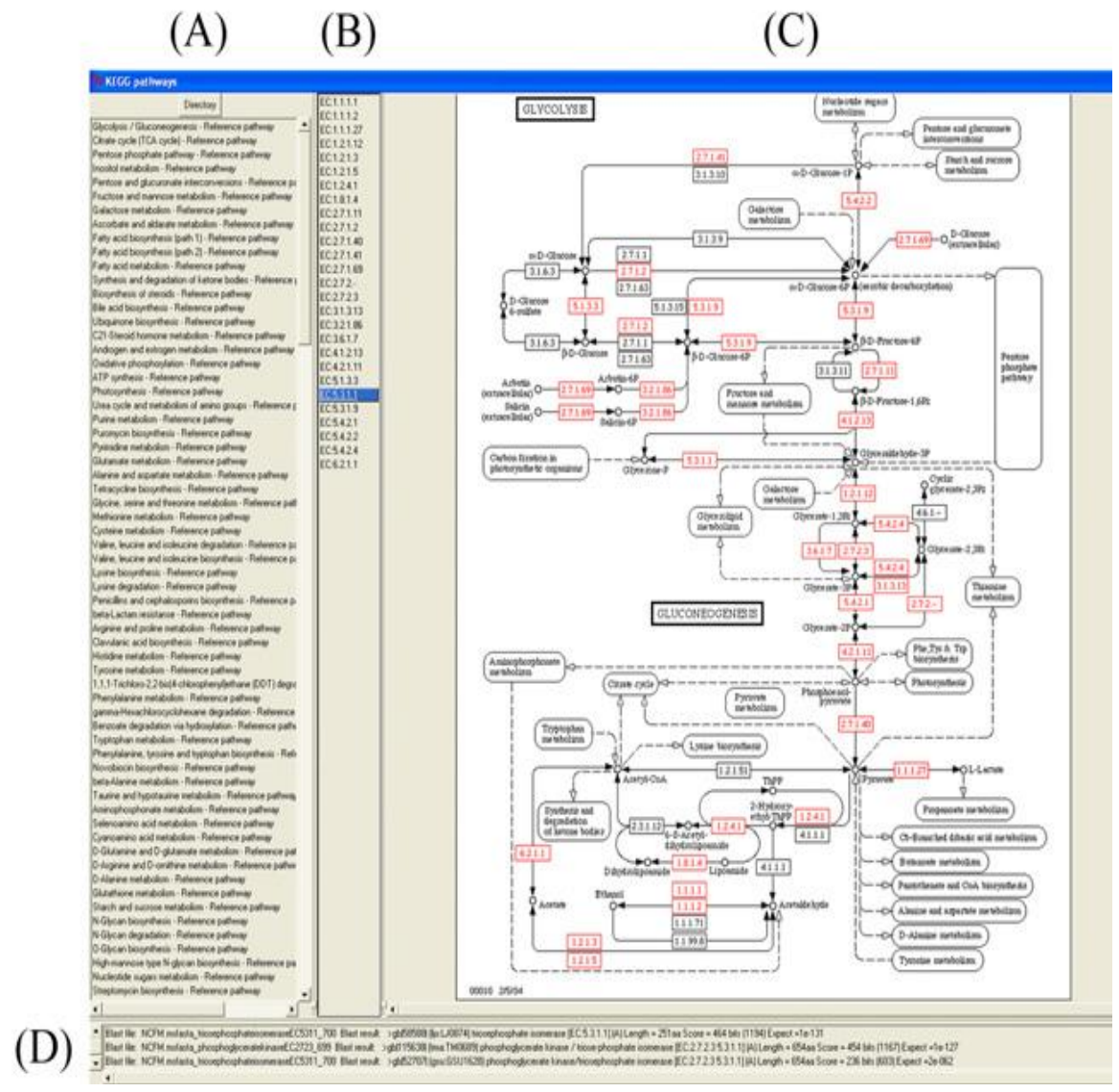

Figure 5: Interactive KEGG Pathway display. The screenshot illustrates KEGG pathway mapping for the glycolysis / gluconeogenesis pathway using the predicted ORFeome of the GAMOLA annotated *L. acidophilus* NCFM genome as query template [19].

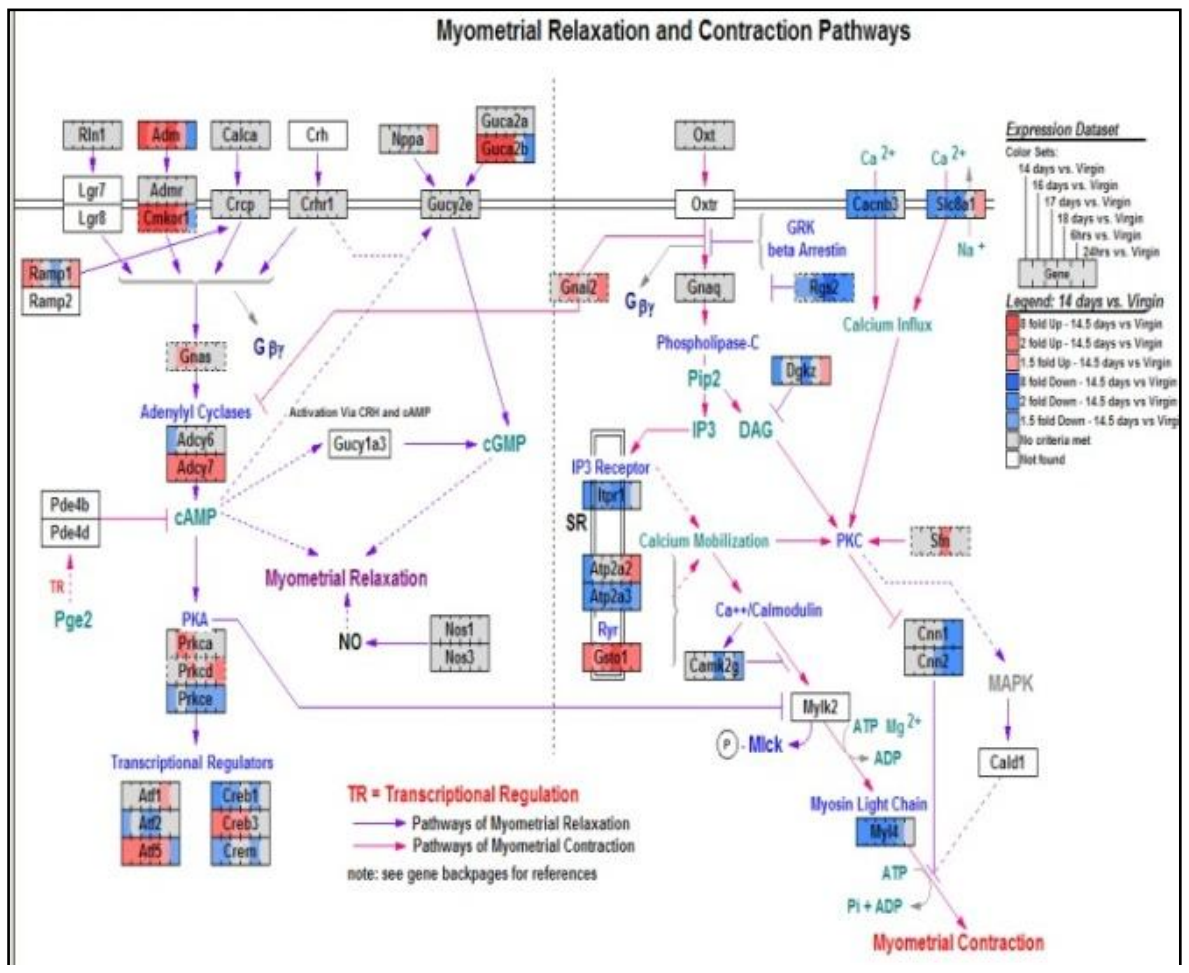

Figure 6: Myometrial Relaxation and contraction Pathway, Image generated by GenMapp.

A metabolic network is a reliable source of information. The reconstruction of GRNs is largely promoted by advances in high-throughput technologies, which enable to measure the global response of a biological system to specific interventions. For instance, large-scale gene expression monitoring using DNA microarrays is a popular technique for measuring the abundance of mRNAs. However, integration of different types of 'omics' data from genomics, proteomics and metabolomic studies can be undertaken. Although the metabolic network has important features for drug discovery, its use in case of human is very limited [18]. Further, Proteomics may yield crucial information on the regulation of biological functions and the mechanism of diseases. In this sense it is a highly promising area for drug discovery. Hence, additional efforts will be required for metabolic network reconstruction and analysis.

## IV. CONCLUSION

From the above tools it can be concluded that metabolic pathways stored as directed acyclic graphs can be considered a basic concept for the visualization tool for metabolic pathway. With respect to visualization, single network views provide little more than brief glimpses of the large datasets. Visualization tools need to support many different types of views, each network view at a different level of detail. Dynamic navigation from one view to another will be a key to showing the connection between different views. Navigating from one time series point to another, for instance, could involve a view showing only the differences between the two time points. If the time points are consecutive, the number of differences will tend to be quite small. A similar approach could be applied to sub-cellular localization information as well. To adequately address each of these issues, active cooperation is required between a variety research fields including graph drawing, information visualization, network analysis and of course biology. Though all the mentioned tool differs significantly in its approach for pathway reconstructions. Hence for future a tool is needed which describe all pathways that make up a cell and how they interact as a system in the overall physiology of an organism is required.

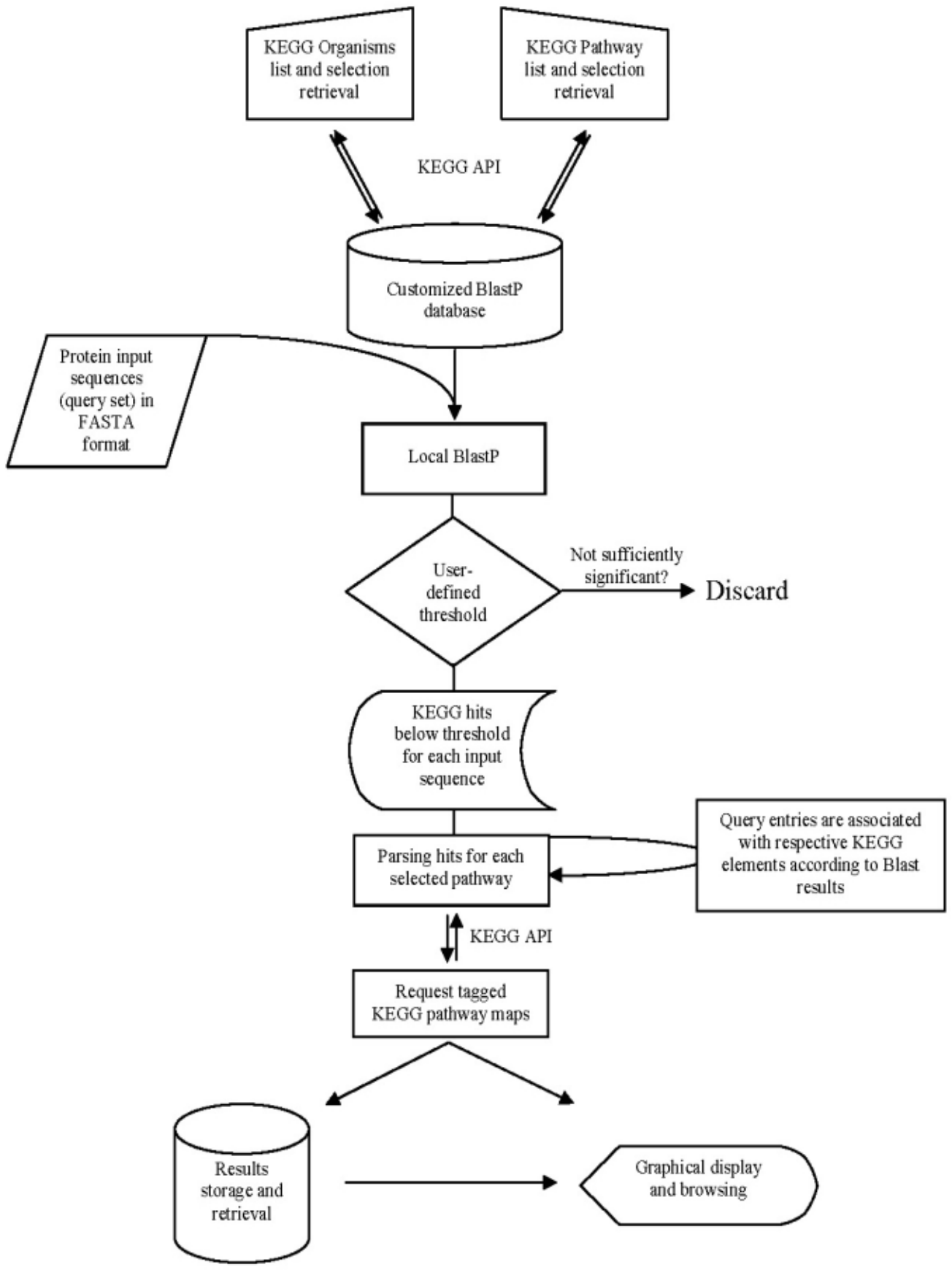

Figure 7: Pathway voyager mapping procedure [17].

## AUTHORS PROFILE

**Ms. Rashmi Rameshwari** has received two master’s degree from one from T.M.B.U., Bhagalpur, Bihar and other from Jamia Hamdard, New Delhi in the area of biotechnology and bioinformatics respectively. She is currently associated with Manav Rachna International University as Assistant Professor in Dept. of Biotechnology. Her research interests include systems biology, proteomics, Microarray Technology, Chemoinformatics, etc.

**Dr. T. V. Prasad** received his master’s degree in Computer Science from Nagarjuna University, AP India and doctoral degree from Jamia Milia Islamia University, New Delhi, India. With over 16 years of academic and professional experience, he has a deep interest in planning and executing major IT projects, pursuing research interest in CS/IT and bioinformatics. He has authored over 65 publications in reputed journals and conferences. He has also authored 4 books. He has also held respectable positions such as Deputy Director with Bureau of Indian Standards, New Delhi. His areas of interest include bioinformatics, artificial intelligence, consciousness studies, computer organization and architecture. He is a member of reputed bodies like Indian Society of Remote Sensing, Computer Society of India, APBioNet, International Association of Engineers, etc.

Table 1
**COMPARATIVE STATEMENT OF VARIOUS NETWORK VISUALIZATION AND ANALYSIS SOFTWARE**

| S. No. | Parameters/ Features | Ingenuity Pathway Analysis | Metacore | Pathway Studio | GenMAPP | Cytoscape | Pathway Voyager | PathFinder |
|---|---|---|---|---|---|---|---|---|
| 1 | Developed by | Ingenuity Systems Inc. | GeneGo Inc. | Ariadne Genomics Inc. | Gladstone Institutes | Paul Shanon, Andrew Markiel et al. | Eric Altermann and Todd R Klaenhammer | Alexander Goesmann et.al |
| 2 | Description | Database of Biological networks created from millions of relationships, between proteins, genes ,complexes, cells, tissues, drugs and diseases | A manually curetted database of human Protein-protein interaction & Protein-DNA interactions, transcriptional factors, Signaling, metabolism and bioactive molecules | These databases are a collection of eukaryotic molecular interactions generated by MedScan Text to knowledge suit using the entire PubMed database and 43 full text journals. Also works with public database of signaling and biochemical pathways | Archived Maps were drawn based on textbooks, articles and public pathway databases or generated from the public database maintained by the Gene Ontology Project | Software for integrating biomolecular interaction Networks with high-throughput expression data and other molecular states into a unified conceptual framework | Utilizes the KEGG online database for pathway mapping of partial and whole prokaryotic genomes | A tool for the dynamic visualization of metabolic pathways based on annotation data. |
| 3 | Availability | Commercial | Commercial | Commercial | Public | Public | Public | Public |
| 4 | Based on database | Ingenuity pathways knowledge base | Human Database | Mammalian Database: ResNet 5, ResNet plant database, KEGG, BIND, HPRD | KEGG | KEGG | KEGG | RDBMS is usec along with KEGG |
| 5 | Web access | Enabled | Enabled | Enabled | Enabled | Enabled | Enabled | Enabled |
| 6 | Platform | Java 1.5 or higher | Java | Python | Java | Java | Perl/TK | Perl/TK |
| 7 | Special features | Solution is given to Pharmaceutical, Bio-technology, and Academics | unique ability to concurrently visualize multiple types of experimental data such as gene expression, proteomic, metabolomics, SAGE, MPSS, SNP, HCS, HTS, microRNA and clinical and pre-clinical phenotypic data | Analyze proteomic, Metabolomics and other high throughput data. | has graphics tools for constructing and modifying pathways. Used for analyzing microarray data include statistical filters and pattern finding algorithms such as hierarchical clustering | mRNA Expression profiles, Gene annotations from Gene Ontology (GO) & KEGG. Incorporates statistical analysis | No dedicated hardware or software are necessary to analyze given datasets | Aim at comparing pathways at microscopic level and therefore it can be used for dynamic visualizations of metabolisms from a whole genome perspective |
| 8 | Drawbacks | Limited to Human, Mouse, Rat and Canine | Specific server is required | It offers a wizard interface for creating very simple network and data queries and only Biological Networks provides a language interface for expressing such queries. | Generally used to explore Microarray data | Focuses only on high-level networks, low-level models of components and interactions addressed by ongoing projects such as Ecell (Tomita et al. 1999), VirtualCell; mechanisms for bridging high-level interactions with lower level, physico-chemical | For certain selectable pathways (e.g. Ribosomal reference pathway) KEGG does not yet support organism independent marking. For practical reasons, no hits will be displayed for these pathways | Uses RDBMS based internet application. Being integrated into the locally developed genome annotation system GENDB with extended functionality |

| S. No. | Parameters/ Features | Ingenuity Pathway Analysis | Metacore | Pathway Studio | GenMAPP | Cytoscape | Pathway Voyager | PathFinder |
|---|---|---|---|---|---|---|---|---|
| | | | | | | models of specific biological processes are required | | |
| 9 | Integrated with | Only based on Ingenuity Product | Gene Go | MedScan | MAPP | Numerous | GAMOLA | GENDB |
| 10 | Storing result | Present | Present | Present | Present | Present | Present | Present |
| 11 | Data format | Not Specific. | Not Specific | ResNet Exchange XML formats. | CSV, GPML, WikiPathways , MAPP | PSI-MI, GML/XGMML | Fasta Files | EMBL OR Genbank |
| 12 | Graph comparison for species | Present | Present | Present | Present | Present | Present | Present |
| 13 | Graphical user Interface | Present | Present | Present | Present | Present | Present | Present |
| 14 | Visualization Technique | Present | Present | Present | Present | Present | Present | Present |
| 15 | Ease of use and report generation | Excellent | Excellent | Excellent | Excellent | Very Good | Good | Good |
| 16 | Graphical Representation | Present | Present | Present | Present | Present | Present | Present |
| 17 | Classification Technique | Compare affected pathways and phenotypes across time, dose, or patient population. | Disease, Tissue, Species, Sub cellular localization, Interactions, Metabolites | Interrogate Different species, Multiple genomes | Hierarchical clustering | None | Hierarchical clustering | Chunks and Subway |
| 18 | Web browser | Internet Explorer 6.0 or higher | Internet Explorer 6.0 or higher | Internet Explorer 6.0 or higher | Internet Explorer 6.0 or higher | Internet Explorer 6.0 or higher | Internet Explorer 6.0 or higher | Internet Explorer 6.0 or higher |
| 19 | Memory | 512 Mb (minimum), 1GB (Recommended) | 4 GB | 1 GB | 512 Mb | 512 Mb | 512 Mb | 512 Mb |
| 20 | Operating System supported | Vista, Window XP and Macintosh 10.3,10.4, 10.5 | Linux 2.1, 3.0 or Red Hat 9.0 | Windows | Windows | Window | Windows | UNIX, Windows |
| 21 | Reference URL | www.ingenuity.com/ | www.genego.com | www.ariadnegenamics.com | www.genmapp.org/ | www.cytoscape.org | www.bioinformatics.ai.sri.com/ptools | http://bibiserv.TechFak.UniBielefeld.DE/pathfinder/ |
| 22 | References | [12] | [32] | [24] | [14] | [15] | [17] | [25], [26] |